\documentclass[aps,prl,article,twocolumn,showpacs,preprintnumbers,amsmath,amssymb,superscriptaddress]{revtex4}

\date{\today}
\usepackage{epsfig}
\usepackage{subfigure}
\usepackage{graphicx}
\usepackage{dcolumn}
\usepackage{bm}
\usepackage[colorlinks,linkcolor=blue,hyperindex,CJKbookmarks]{hyperref}
\usepackage{float}
\usepackage{hyperref}
\usepackage{comment}
\begin{document}

\title{Using Nonequilibrium Dynamics to Probe Competing Orders in a Mott-Peierls System}
\author{Y. Wang }
 \affiliation{Department of Applied Physics, Stanford University, California 94305, USA}
 \affiliation{SLAC National Accelerator Laboratory, Stanford Institute for Materials and Energy Sciences, 2575 Sand Hill Road,
Menlo Park, California 94025, USA}
\author{B. Moritz}
\affiliation{SLAC National Accelerator Laboratory, Stanford Institute for Materials and Energy Sciences, 2575 Sand Hill Road,
Menlo Park, California 94025, USA}%
\affiliation{Department of Physics and Astrophysics, University of North Dakota, Grand Forks, North Dakota 58202, USA}
\date{\today}
\author{C.-C. Chen}
\affiliation{Advanced Photon Source, Argonne National Laboratory, Argonne, Illinois 60439, USA}%
\author{C. J. Jia}%
\affiliation{SLAC National Accelerator Laboratory, Stanford Institute for Materials and Energy Sciences, 2575 Sand Hill Road,
Menlo Park, California 94025, USA}
\author{M. van Veenendaal}
\affiliation{Advanced Photon Source, Argonne National Laboratory, Argonne, Illinois 60439, USA}%
\affiliation{Department of Physics, Northern Illinois University, De Kalb, Illinois 60115, USA}
\author{T. P. Devereaux}
\affiliation{SLAC National Accelerator Laboratory, Stanford Institute for Materials and Energy Sciences, 2575 Sand Hill Road,
Menlo Park, California 94025, USA}%
\affiliation{Geballe Laboratory for Advanced Materials, Stanford University, California 94305, USA}
\begin{abstract}
Competition between ordered phases, and their associated phase transitions, are significant in the study of strongly correlated systems. Here we examine one aspect, the nonequilibrium dynamics of a photoexcited Mott-Peierls system, using an effective Peierls-Hubbard model and exact diagonalization. Near a transition where spin and charge become strongly intertwined, we observe anti-phase dynamics and a coupling-strength-dependent suppression or enhancement in the static structure factors. The renormalized bosonic excitations coupled to a particular photoexcited electron can be extracted, which provides an approach for characterizing the underlying bosonic modes. The results from this analysis for different electronic momenta show an uneven softening due to a stronger coupling near $k_F$. This behavior reflects the strong link between the fermionic momenta, the coupling vertices, and ultimately the bosonic susceptibilities when multiple phases compete for the ground state of the system.
\end{abstract}

\maketitle

Emergent phenomena in strongly-correlated materials arise from the interplay between degrees of freedom, resulting in a delicate balance of electron, spin, lattice, and orbital interactions. This is most clearly displayed in the relationship between excitations around competing ground states.
For example, in high-temperature superconductors such as the cuprates \cite{abbamonte2005spatially, ghiringhelli2012long, chang2012direct, keimer2015quantum} and iron-pnictides \cite{chuang2010nematic, paglione2010high, johnston2010puzzle, chu2012divergent,fernandes2014drives},
antiferromagnetism (AFM), charge-density-wave (CDW), and nematic order are believed to be the key competitors of superconductivity (SC) \cite{kivelson2003detect, fujita2011progress, davis2013concepts, PhysRevB.92.024503}.
New bosonic excitations induced by quantum fluctuations in the vicinity of quantum phase transitions (QPT) could potentially mediate SC \cite{orenstein2000advances, sidorov2002superconductivity, chubukov2003spin}.
In correlated materials displaying both AFM and CDW orders, the spin and charge excitations are intimately linked and may behave quite differently when either order is dominant.
This interplay is expected to become profound near a QPT where neither order is dominant, $i.e.$ in the crossover between two phases, and they compete for the underlying ground state \cite{sachdev2000quantum, RevModPhys.73.797, coleman2005quantum, gegenwart2008quantum, coldea2010quantum}.

Despite the importance of this emergent behavior, surprisingly little is known about the evolution of excitations in such systems. Nonequilibrium studies provide the ability to separate and track various competing orders and offers an effective approach for characterizing and analyzing their dynamics \cite{fausti2011light, lee2012phase, hellmann2012time, kim2012ultrafast, dal2014snapshots, wang2014real, gerber2015direct, gorshkov2011tunable, rubbo2011resonantly, carrasquilla2013scaling, hazzard2013far, schachenmayer2013entanglement}. 
As a powerful and widely used tool, an ultrafast pump allows for the photomanipulation of the delicate balance between different competing orders \cite{rini2007control, Yonemitsu20081, PhysRevLett.102.106405, tomeljak2009dynamics, schmitt2008transient,  lu2012enhanced, PhysRevB.88.045107,rincon2014photoexcitation, lu2015photoinduced}. 
To understand how electrons dressed by bosonic excitations form quasiparticles in systems with intertwined orders, angle-resolved photoemission (ARPES) in the time-domain can give quantitative insight into the integrity of the quasiparticle as well as its renormalized dispersion \cite{Damascelli:2003kq,zhou2007angle, berciu2006green}. 
However, one must still infer which bosonic excitations give rise to renormalization, embodied solely in the single-particle self energy. 
In contrast, those bosonic excitations are directly visible via inelastic x-ray \cite{abbamonte2005spatially, ghiringhelli2012long, chang2012direct}, neutron \cite{fong1995phonon, weber2009signature}, and Raman scattering \cite{devereaux2007inelastic, gupta2006raman} as well as other optical methods \cite{basov2005electrodynamics, basov2011electrodynamics}, yet difficult to correlate back with the properties of the renormalized electron measured via ARPES.

To link the two descriptions and monitor nonequilibrium dynamics from the perspective of both the electronic and bosonic degrees of freedom, in this study we simulate the nonequilibrium dynamics of a Peierls insulator upon instantaneous photoexcitation.
Whereas ARPES provides the spectral function, which has been well characterized in previous studies \cite{Damascelli:2003kq,zhou2007angle}, we instead correlate the spectra with a detailed measurement of the charge and spin bosonic excitations coupled to the excited electron, unveiling the link between the fermionic and bosonic renormalizations in a system having intertwined spin and charge orders.

\begin{figure}[!t]
\centering
\includegraphics[width=8.5cm]{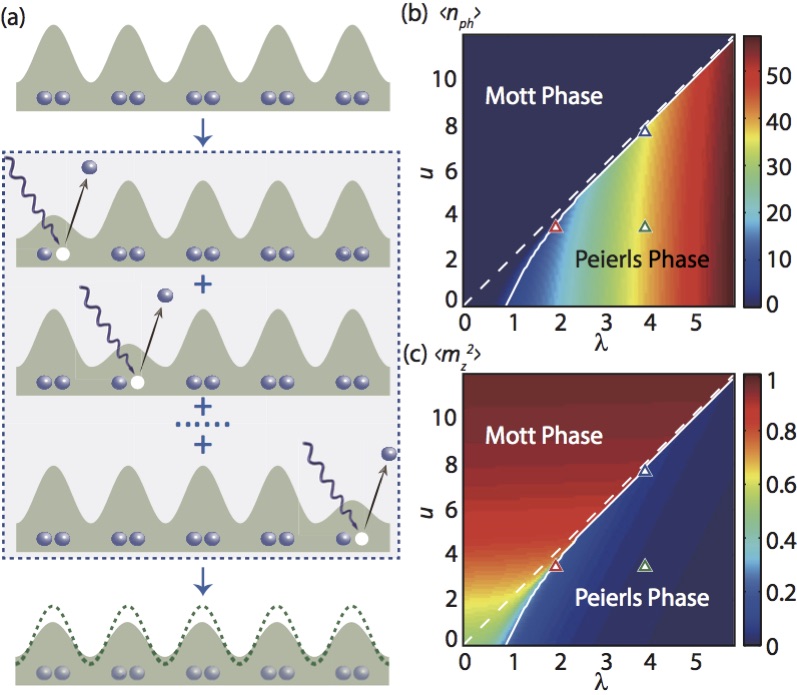}
\caption{\label{fig:1} 
(a) Diagram showing the photoexcitation process from the Peierls ground state (top panel). This process is realized by a coherent, uniform removal of an electron with zero net momentum (middle panel). The consequence is projection of the ground state into a Hilbert space with $N\!-\!1$ particles and a reduced charge modulation (bottom panel).
(b) Phonon occupations and (c) local moment at various $\lambda$ and $u$. In both figures, the dashed line denotes the anti-adiabatic limit phase boundary ($u_{\rm{eff}}\!=\!0$) and the solid line denotes the boundary of Peierls phase in terms of ground state degeneracy. The triangles indicate the positions of correspondingly colored lines of Figs.~\ref{fig:2} (b) and (c).  }
\end{figure}

Utilizing an effective model that captures CDW/AFM competition, we first create a momentum-resolved photohole resulting from photoexcitation of an electron, and then directly calculate the charge and spin excitations of the remnant system [see Fig.~\ref{fig:1}(a)].
By comparing the dynamics across parameter space, we find that the frequency associated with the charge and spin responses reflects the bosonic excitations associated with the photoexcited electron.
This link between bosonic excitations and the fermionic momentum of the photoexcited electron is reflected in the uneven softening of modes approaching the phase boundary.


The physics of the Mott-Peierls system can be captured in the Hubbard-Holstein model (HHM)\cite{Hubbard,Holstein}, which has been well studied in equilibrium: the presence of both electron-electron (\textit{e-e}) and electron-phonon (\textit{e-ph}) effects leads to CDW/AFM competition and a metallic region between the ordered phases\cite{clay2005intermediate, fehske2008metallicity, bauer2010competing,nowadnick2012competition, murakami2013ordered, hohenadler2013excitation, Greitemann2015arxiv}. In this model $\pi$ momentum dominates the underlying order for both spin and charge; therefore, we introduce the Peierls-Hubbard model (PHM), which simplifies the lattice degrees of freedom to a uniform dimerization $\mathcal{H}_{\rm PHM} = \mathcal{H}_{\textit{e-e}}+ \mathcal{H}_{\textit{e-ph}}$:
\begin{align}\label{HHM}
\mathcal{H}_{\textit{e-e}} &=-t_h\sum_{i,\sigma}(c_{i\sigma}^\dagger c_{i+1,\sigma}+h.c.)+U\sum_i n_{i\uparrow}n_{i\downarrow}\nonumber\\
\mathcal{H}_{\textit{e-ph}}&=-\frac{g}{\sqrt{N}}(b^\dagger +b)\sum_{i,\sigma}(-1)^in_{i\sigma}+\Omega\, b^\dagger b,
\end{align}
where $t_h$ is the nearest-neighbor hopping integral, $c_{i\sigma}^\dagger$ ($c_{i\sigma}$) and $n_{i\sigma}$ are the electron creation (annihilation) and number operator at site $i$ of spin $\sigma$, $U$ is the on-site Coulomb repulsion, and $b^\dagger$ ($b$) and $\Omega$ are the phonon creation (annihilation) operator and frequency, respectively.
The dimensionless \textit{e-e} and \textit{e-ph} coupling strengths are defined as $u=U/t_h$ and $\lambda=g^2/t_h\Omega$, respectively. The phonon frequency is set to $\Omega\!=\!t_h$ as in Ref.~\cite{clay2005intermediate}. 
The calculations are performed on one-dimensional chains of $N\!=\!10$ sites with periodic boundary conditions and maximum phonon occupation $M\!=\!127$. 
We use the parallel Arnoldi method\cite{lehoucq1998arpack} to determine the ground state wavefunction and the Krylov subspace technique\cite{manmana2007strongly, balzer2012krylov, park1986unitary, hochbruck1997krylov, moler2003nineteen, SUPMAT} to evaluate the evolution of a state $|\psi(t\!+\!\delta t)\rangle\!=\!e^{-iH\delta t}|\psi(t)\rangle$. 

Based on this PHM Hamiltonian, Figs.~\ref{fig:1}(b) and (c) show the phonon occupation and local moment at half-filling as function of $u$ and $\lambda$ in equilibrium obtained from our ED calculation. The dashed line indicates the phase boundary in the anti-adiabatic limit where $u_\textrm{eff}\!=\!0$, while the solid line tracks the numerical boundary where the translational symmetry breaks and the ground state changes from a doubly-degenerate (Peierls phase) to non-degenerate.
This boundary approaches the anti-adiabatic line asymptotically in the strong-coupling limit, while the metallic phase broadens as electron itinerancy dominates in the intermediate and weak-coupling regime. This equilibrium phase diagram is consistent with results from the HHM\cite{clay2005intermediate,bauer2010competing,fehske2008metallicity,nowadnick2012competition, murakami2013ordered, hohenadler2013excitation, Greitemann2015arxiv}, demonstrating the effectiveness of our model in capturing competing CDW/AFM orders [see the Supplementary Material\cite{SUPMAT} for detailed discussions].

\begin{figure*}[!ht]
\begin{center}
\includegraphics[width=18cm]{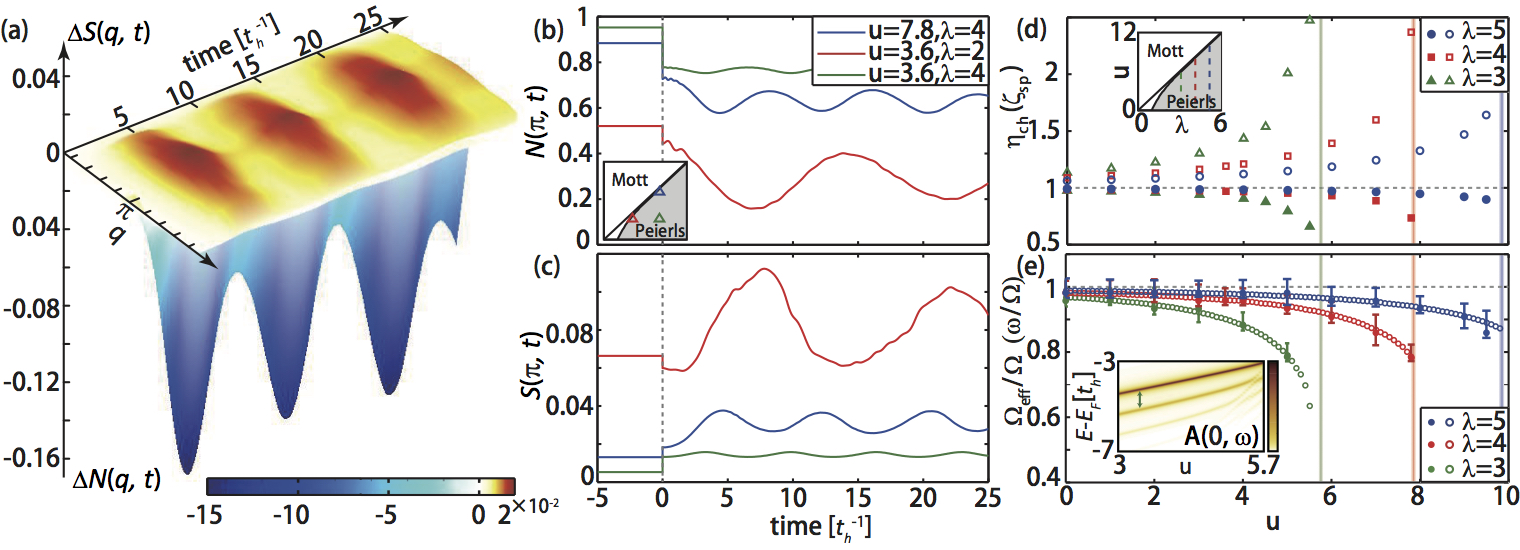}
\caption{\label{fig:2} (a) False color plots of the spin density $\Delta S(q,t)$ and charge density $\Delta N(q,t)$ dynamics for $u\!=\!7.8$ and $\lambda\!=\!4$ (strong-coupling regime) after photoexcitation at the $\Gamma$ point. (b,c) Evolution of (b) $N(\pi,t)$ and (c) $S(\pi,t)$ with various parameters (shown in the inset).
(d) The suppression of CDW $\eta_{ch}$ (solid markers) and enhancement of AFM $\zeta_{sp}$ (open markers) in the Peierls phase (shown in the inset).
(e) The comparison of dominant energy scales in $N(\pi,t)$ and $S(\pi,t)$ dynamics (solid circles) and the renormalized phonon frequencies (open circles) evaluated via the equilibrium spectral function $A(k,\omega)$, as indicated in the inset. The frequencies are compared along the $u$ axis with $\lambda=3$, 4 and 5. The error bars denote the corresponding half width at full maximum in the Fourier spectrum due to a finite time window. 
}
\end{center}
\end{figure*}

Aided by the phase diagram in Fig.~\ref{fig:1}(b), we study the dynamics upon photoexcitation of an electron $|\psi(0_+)\rangle \!=\! c_{k\sigma}|G\rangle$ from the ground state $|G\rangle$ in the Peierls phase
We first concentrate on zero momenta where $c_{k\sigma}\!=\!\frac1{\sqrt{N}}\sum_{j\sigma}c_{j\sigma}$ [Fig.~\ref{fig:1}(a)].
Instead of focusing on the photoexcited electron with links to ARPES, we analyze the temporal dynamics of charge  and spin in the remnant system by evaluating the instantaneous structure factors $N(q,t)\!=\!\langle \psi (t)| \rho^{(c)}_{-q}\rho^{(c)}_{q}|\psi(t)\rangle$ and $S(q,t)\!=\!\langle \psi(t) | \rho^{(s)}_{-q}\rho^{(s)}_{q}|\psi(t)\rangle$,
where $\rho^{(c/s)}_q\!=\!\sum_k \left(c^\dagger_{k+q\uparrow}c_{k\uparrow}\pm c^\dagger_{k+q\downarrow}c_{k\downarrow}\right)$.
These nonequilibrium structure factors after photoexcitation do not reflect the equilibrium properties of simply photodoping, but reveal the information about competing order and the underlying bosonic excitations.

To provide a global perspective on the dynamics in terms of time and momentum, we first focus on a strong-coupling set of parameters near the phase boundary [$u\!=\!7.8, \lambda\!=\!4$, indicated by the blue triangle in Figs.~1(b) and (c)].
Before photoexcitation, the charge correlation peaks sharply at $q\!=\!\pi$ due to the CDW order, while spin is weak and broad\cite{SUPMAT}. 
Although different in equilibrium, their dynamics are tightly related.
Fig.~\ref{fig:2}(a) shows the evolution of the charge [$\Delta N(q,t)\!=\!N(q,t)\!-\!N(q,0_+)$] and spin structure factors [$\Delta S(q,t)\!=\!S(q,t)\!-\!S(q,0_+)$] after photoexcitation. Their momentum distribution (especially for charge) indicates the dynamics can be captured roughly by the time structure at $q\!=\!\pi$. In this sense along the time axis, the evolution of the charge and spin structure factors reveals an anti-phase dynamics with the same frequency, reflecting the underlying competition between the two orders, which is manipulated by photoexcitation. This situation occurs only when the underlying orders are intertwined and is expected to be amplified in proximity to a transition between the two phases as they compete for the ground state.

To further investigate the nonequilibrium dynamics of intertwined orders, we next compare $N(\pi,t)$ and $S(\pi,t)$ for various coupling parameters as shown in Figs.~\ref{fig:2}(b) and (c), respectively.
Deep in the Peierls phase (green lines), one sees a rather robust CDW with $N(\pi,t)$ suppressed less than $5\%$ following photoexcitation (it drops immediately at $t\!=\!0$ due to photodoping), as a result of strong localization without significant competition. However, a strong suppression/enhancement of the charge/spin structure factors occurs near the crossover (blue and red lines). 
To parameterize the robustness against photoexcitation, we define the coefficients of charge suppression $\eta_{ch}\!=\!\min_{t>0} N(\pi,t)/N(\pi,0_+)$ and spin enhancement $\zeta_{sp}\!=\!\max_{t>0} S(\pi,t)/S(\pi,0_+)$.
As shown in Fig.~\ref{fig:2}(d), $\eta_{ch}$/$\zeta_{sp}$ decreases/increases rapidly towards the phase boundary as the charge and spin excitations become increasingly intertwined, while both asymptotically approach $1$ when moving away as the CDW becomes more robust. 

\begin{figure*}[!ht]
\begin{center}
\includegraphics[height=6cm]{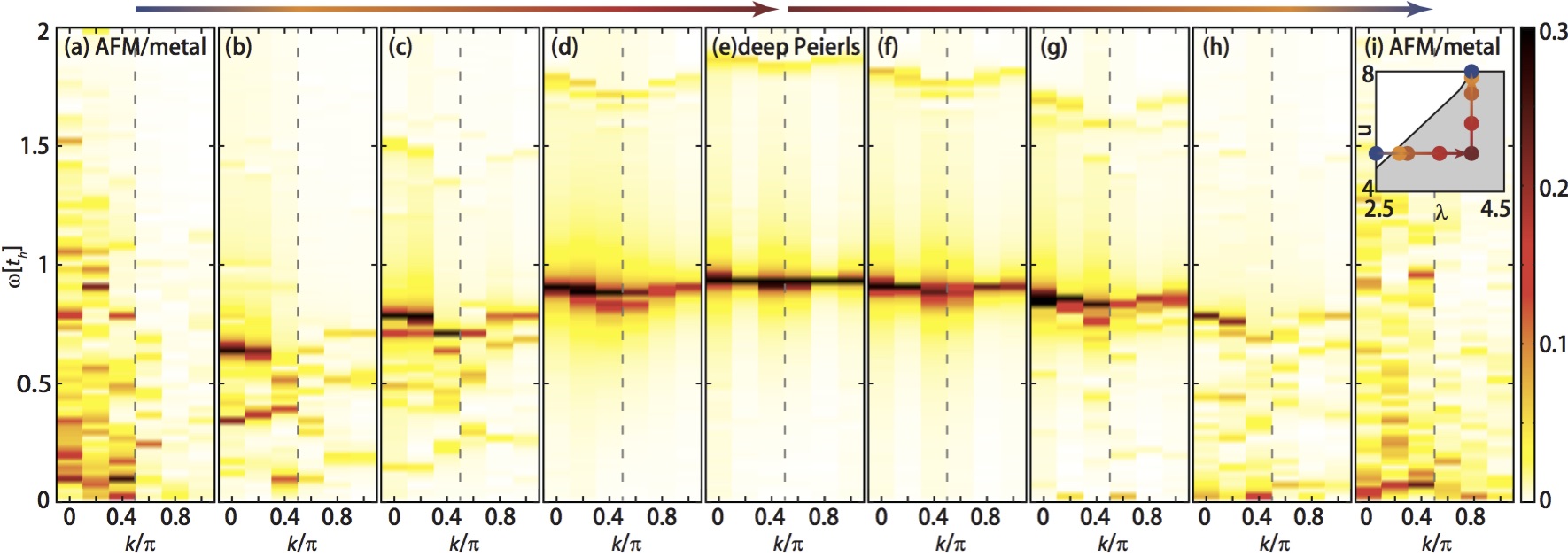}
\caption{\label{fig:3}
The Fourier spectrum of $N(\pi,t)$ for photoexcitation with different fermion momentum, calculated along a path through the phase diagram indicated by the colored arrow. (a-e) follows the path with fixed $u=5$ and increasing $\lambda$; while (e-i) follows the path with fixed $\lambda=4$ and increasing $u$. Among them, (e) is deep in the Peierls phase; (b) and (h) are close to the phase boundary; (a) and (i) are outside the phase. $k_F$ is denoted by the dashed lines. The inset reflects the path in parameter space.
}
\end{center}
\end{figure*}

We further analyze the periodicity of the structure factors over a much longer time window, which is crucial for classifying the low-energy bosonic modes as well as the origin of competing orders\cite{hellmann2012time}. We calculate the (average shifted) Fourier spectrum over a time window from 0$\sim$256$\,t_h^{-1}$ and extract the dominant frequency $\omega$ [see Fig.S1 in the Supplemental Material\cite{SUPMAT}]. In contrast to the softened phonon mode at $\omega=0$ associated with the structural deformation inside the Peierls phase\cite{weber2015phonon, nowadnick2015renormalization}, this frequency reflects electron dynamics relative to the deformation.
At the same time, the renormalized phonon frequency $\Omega_{\rm eff}$, as a result of interactions in the vicinity of the crossover\cite{CDWtheory, weber2015phonon,nowadnick2015renormalization}, can be evaluated through the single-particle spectral function $A(k,\omega)$ which reflects the dynamics of the photohole. As shown in Fig.~\ref{fig:2}(e), the agreement between $\omega$ and $\Omega_{\rm eff}$ for various parameters indicates the bosonic excitations associated with the original photoexcited electron can be revealed by the dynamics in the remnant system, connecting the two descriptions. Physically, this oscillatory dynamics represent an energy transfer between the lattice and the electrons, which has been observed in many experiments\cite{schmitt2008transient, kim2012ultrafast, hellmann2012time}.

We see that the spin and charge response in a photoexcited, nonequilibrium system reveals information about the basic underlying bosonic excitations.
Here they are dominated by the bare phonon deep in the Peierls phase, but become increasingly renormalized close to the crossover where spin, charge and phonon are intimately intertwined. This can be especially significant in those complex systems where the bosonic mode cannot be directly measured or easily distinguished in the ARPES spectra. 
In addition to the bosonic excitations near phase boundary for a zero-momentum (or $\Gamma$ point) photoexcitation, a natural follow-up question is whether all the electrons ``feel'' the same intertwined bosonic excitations, particularly near the phase boundary where the competition between the Mott and Peierls physics tends to be delicately balanced.

To answer the above question, we further examine the dynamics associated with photoexcitation for various fermionic momenta $k$ (rather than the restricted $k=0$) as shown in Fig.~\ref{fig:3}. Deep in the Peierls phase [Fig.~\ref{fig:3}(e)] where the bosonic excitations are gapped, the dynamics appear uniform in momentum: all electrons feel a single dispersionless bare phonon as discussed previously. However, approaching the crossover [Figs.~\ref{fig:3}(b-d) and (f-h)], the bosonic excitations show a continuous, but uneven, softening as a function of $k$, which results from a renormalization by the electronic susceptibility.
Due to stronger coupling to the electrons near the Fermi surface, the bosonic mode energy softens faster for $k\!\sim\!k_F$. At the same time, a large number of low-energy excitations appear, especially around $k_F$ [see Fig.~\ref{fig:3}(h)]. Once outside the Peierls phase [Fig.~\ref{fig:3}(a) and (i)], the bare phonon frequency is no longer visible and the bosonic spectra display a low-energy continuum as spin excitations become gapless in the Mott phase.
This is the nature when crossing between the two phases -- even though both $U$ and $\lambda$ may be large, the bosonic excitations lie at low energies and are strongly entwined with a photohole having momentum near $k_F$. 
 
In this sense, the photoexcited dynamics not only reflect the bosonic mode coupled to the electrons, but also reveal the softening of bosonic excitations due to intertwined orders near the phase boundary.
More importantly, this unifies our understanding of fermionic and bosonic coupling: the spin and charge susceptibility $\chi(q,\omega)$ reflects the underlying physics, but also must be linked to additional information about the fermionic momentum and energy. While the coupling parameter $g_q\!=\!g\delta_{q,\pi}$ in Eq.~\ref{HHM} is $k$-independent, the effective bosonic spectra extracted from the dynamics associated with photoexcitation tell a deeper story. In the regime when either order is dominant, the dynamics faithfully reflect the spin and charge susceptibility, with little dependence upon the fermionic momenta of the photoexcited electrons; 
however, the observed dynamics turn out to be highly dependent on the fermionic momentum near the crossover. In other words, not all the electrons are sensitive to the same bosonic excitations in a strongly intertwined system with competing orders. This means that the coupling vertex, which would show up in the single-particle self energy, renormalizing fermions by spin and charge excitations, must be strongly momentum-dependent in such a system near a transition due to the combined impact of the $e-e$ and $e-ph$ coupling.

To summarize,we have studied the dynamics of a Mott-Peierls system after photoexcitation from the Peierls phase using an effective Peierls-Hubbard model with exact diagonalization. We found that the suppression of charge and enhancement of spin correlations reflects the underlying competition, which may be further increased near a phase boundary. 
Furthermore, the dominant frequency reveals the bosonic excitation, here renormalized phonons, coupled to the photoexcited electron. By examining the dynamics following photoexcitation of electrons with different momenta, we were able to observe an uneven softening of the modes, renormalized by the electronic susceptibility, and the possible emergence of additional low-energy modes with a more complicated structure near the crossover. 
This momentum dependence can be amplified near a QPT in a system with competing orders. 
Unlike the traditional bosonic spectra which integrate out the fermionic momenta, these nonequilibrium results provide a new perspective in the study of strongly correlated systems with intertwined orders.

We thank H.C.~Jiang, P.~Kirchmann, W.-S.~Lee and Z.-X.~Shen for insightful discussions. This work was supported at SLAC and Stanford University by the US Department of Energy, Office of Basic Energy Sciences, Division of Materials Sciences and Engineering, under contract No.~DE-AC02-76SF00515 and by the Computational Materials and Chemical Sciences Network (CMCSN) under contract No.~DE-SC0007091. Y.W.~also was supported by the Stanford Graduate Fellows in Science and Engineering. C.C.C.~is supported by the Aneesur Rahman Postdoctoral Fellowship at Argonne National Laboratory (ANL), operated by the U.S. Department of Energy (DOE) Contract No.~DE-AC02-06CH11357. M.v.V.~is supported by the DOE Office of Basic Energy Sciences (BES) Award No.~DE-FG02-03ER46097 and the NIU Institute for Nanoscience, Engineering and Technology. A portion of the computational work was performed using the resources of the National Energy Research Scientific Computing Center supported by the US Department of Energy, Office of Science, under contract No.~DE-AC02-05CH11231.

\bibliography{paper}

\end{document}